# The 4MOST facility simulator: instrument and science optimisation

Th. Boller[a] and T. Dwelly[a]

[a]Max-Planck-Institut für extraterrestrische Physik, PSF 1312, 85741 Garching, Germany

**ABSTRACT**

This paper describes the design and implementation of a facility simulator for the 4 metre Multi-Object Spectroscopic Telescope (4MOST) project, a new survey instrument proposed for the ESO VISTA telescope. The 4MOST Facility Simulator (4FS) has several roles, firstly to optimise the design of the instrument, secondly to devise a survey strategy for the wide field design reference surveys that are proposed for 4MOST, and thirdly to verify that 4MOST, as designed, can indeed achieve its primary science goals. We describe the overall structure of the 4FS, together with details of some important 4FS subsystems. We present the initial results from the 4FS which illustrate clearly the value of having a functioning facility simulator very early in the conceptual design phase of this large project.

## 1. INTRODUCTION

The 4FS consists of three different major components, the Operations Simulator (OpSim), the Throughput Simulator (TS), and the Data Quality Control Tools (DQCT), each of which have specified tasks, as described in Section Simulation System Definition. The OpSim component is located at MPE, Garching, the TS at GEPI, Paris, and the DQCT at IoA, Cambridge. Data flow between the major components is carried out over the Internet through an rsync directory system located at MPE, Garching. The 4FS accepts input data in the format of mock catalogues of targets, together with template spectra of targets. Each mock catalogue represents one DRS. The operation of 4FS is controlled through parameter files issued by the Systems Engineer, which define the set-up of each system that should be to be simulated.

## 2. 4FS OVERVIEW

The top-level architecture of the 4FS is illustrated in Figure 1. The main inputs to each run of the simulator are a set of catalogues of input targets, template spectra for those objects, together with a set of parameters which describe the characteristics of the instrument and the survey strategy. The OpSim component interprets these inputs, and simulates the operations of a 5 year sky 4MOST survey, including a simulation of how fibers will be allocated to targets within each Tile. The OpSim outputs to the DQCT a list of all the input targets to which it was possible to allocate fibers during the course of the survey. The TS takes the template spectra, and folds them through a model of the response of the instrument, telescope and sky, to produce realistic realisations of the spectra that will be collected by 4MOST. The DQCT receives the outputs of the OpSim and TS, and evaluates which of the input targets have been observed sufficiently well to meet pre-determined science citeria. The DQCT calculates the overall scientific success of each DRS, expressed as a numerical Figure of Merit (FoM).

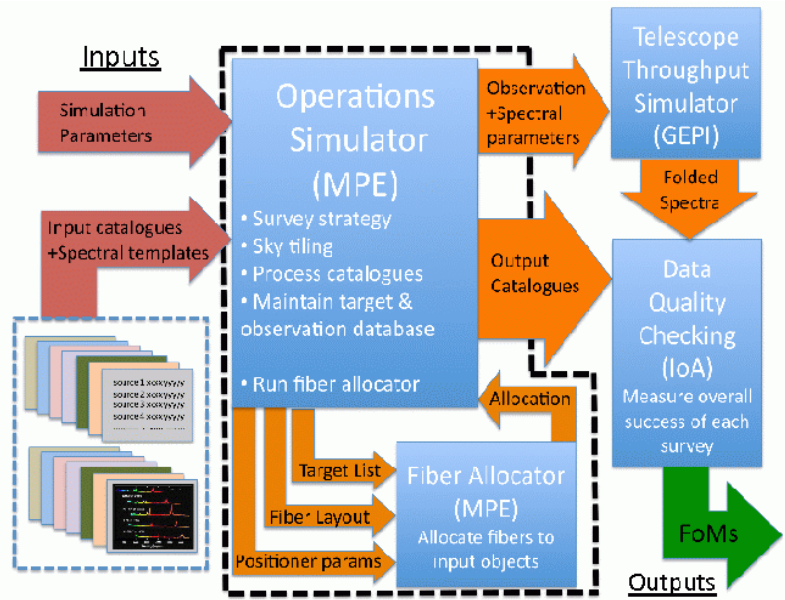

Figure 1: Flow chart illustrating the functional operation of the 4FS, and the data flows between the major components.

## 3. OPERATIONS SIMULATOR INTERNAL STEPS

The OpSim carries out several steps associated with judging the ability of 4MOST, as designed, to carry out the design reference surveys.
1. It receives an input mock target catalogue for each DRS
2. It receives an input parameter file from the Project Office describing the parameter space of the simulation
3. It devises an efficient arrangement of Fields on the sky
4. It simulates the scheduling of a 5-year duration 4MOST sky survey, taking account of moon phase, observing environment (cloud cover, seeing) and sky accessibility.
5. It simulates how an operational 4MOST would assign fibers to targets, taking account of target priorities and requested exposure times, and taking care to avoid collisions between neighbouring fiber positioners.
6. It outputs a list of the input objects that could be observed during a 5-year duration 4MOST survey, including the realized exposure times and environmental parameters.

**3.1 Survey Strategy, Sky Tiling and Observing Conditions**

The OpSim receives as part of its control parameters a description of the desired sky coverage of the final 4MOSTsurvey, including the number of Tiles required in each part of the sky, and the desired exposure time per Tile. Broadly speaking the sky is divided into two regions, the Galactic plane ( $|b| \leq 15$ deg) which will mainly be observed during bright moon phases, and the remainder of the sky which will be observed in dark and grey time. The input coverage request is tuned to match the sky distribution and requested exposure times of the input target catalogues. It is then the task of the OpSim to carry out a simulated 5 year survey that realises a sky coverage that matches the requested sky coverage as closely as possible, taking account of the various environmental observing constraints that would apply in any real 4MOST survey.

The OpSim places hexagonal 4MOST Fields on the sky in an arrangement derived from the icosahedrally symmetric spherical covering patterns of R.H. Hardin et al. [1]. An appropriately spaced pattern is chosen so that gaps between neighbouring fields are reduced whilst also ensuring that the overlapping area is minimised. The choice of tiling pattern is dependent on the FOV of the instrument. The position angles of Fields relative to the Equatorial grid are calculated to reduce gaps between Fields. Future modifications will allow dithered offsets between repeated observations of the same Field to improve the evenness of coverage.

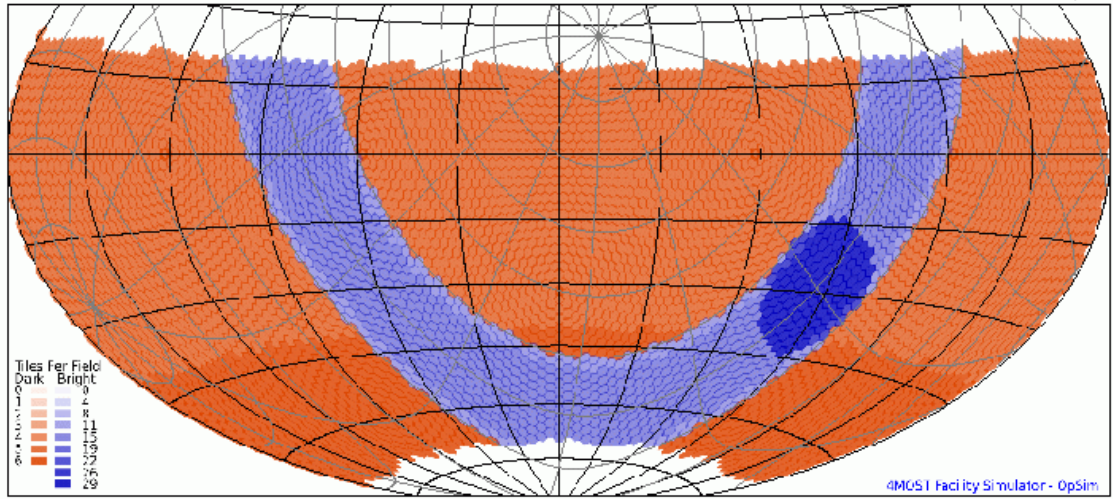

Figure 2. An example of the sky coverage (in Equatorial coordinates) that could be achieved in a 5 year survey with the 4MOST project. The diagram shows the number of 20 minute Tiles observed per Field. Orange areas are observed in Dark time and Blue areas in Bright time. The diagram is valid for a situation where 4MOST is sited on the VISTA telescope with an instrument FOV of 6.1 deg$^2$ (the Goal configuration). Extra time is expended for observations of the Galactic bulge.

The OpSim will simulate the following aspects of the observing conditions expected during 4MOST survey operations.

1. A representation of twilight-to-twilight dark time available per night, and the accessible sky area, as a function of time of year. This calculation will take account of the differences in longitude, latitude and altitude of the two potential sites (La Silla or Paranal) and will be achieved using publicly available ephemeris tools.
2. A representation of the Moon phase and distance for each observed tile. This will be calculated using publicly available ephemeris tools.
3. A randomized assignment of the seeing FWHM for each tile, following the long term distributions measured at La Silla/Paranal. The seeing distribution will be represented either as a tabulated distribution or as an analytic approximation to the observed distribution.
4. A randomized assignment of cloud cover for each night, based on past cloud cover statistics, including an accounting for night time completely lost due to bad weather.

### 3.2. Fiber Positioners

Two different positioner concepts have been considered for 4MOST, named "MuPoz", and "PotzPoz". MuPoz is a dual-rotation axis type positioner, similar in some functional aspects to the positioner technology operation on the LAMOST telescope. PotzPoz is a novel rotation+translation concept. The positioners are arranged in the focal plane in a regular hexagonal grid. Neighbouring positioners can potentially collide, because in order to acheive full coverage of the focal plane the positioners' maximum patrol radii are larger than half the positioner-positioner spacing. The OpSim models the 2-dimensional geometry of the positioner outlines as a collection of vectors and circles, see Figure 3. No part of a positioner arm may approach closer than a fixed distance from any part of a neighbouring positioner. Collision tests are carried out by the OpSim as part of the fiber allocation module.

### 3.3 Fiber Allocation Module

This module determines the allocation of the 1500+ fibers to candidate targets during each Tile. The allocation algorithm works from the object perspective, that is, we first create a ranked list of potential targets. Input targets are ranked by several criteria, including science priority, their requested exposure time and if they have been observed in previous Tiles. We then work down the ranked target list allocating a fiber to each target if possible until we either run out of targets or available fibers. Before allocating a fiber to a target we verify that this allocation would not cause a collision between positioner arms. A pre-determined fraction of fibers are reserved for measurements of the sky background. Each field will in general be observed with several fiber configurations, and the total SNR for a target will typically be built up over several exposures.

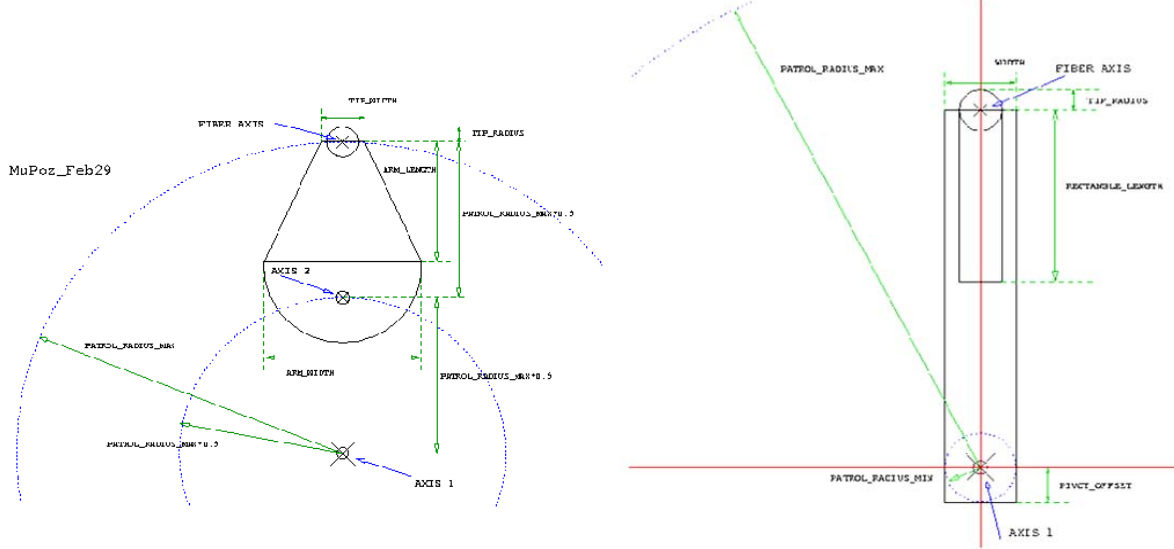

Figure 3: The representation in code of the 2D perimeters of the MuPoz (left) and PotzPoz (right) positioner concepts.

### 3.4 Fiber allocation results

#### 3.4.1 Fiber density plots

The fiber radial scatter plots, see Figure 4 illustrate the placement of positioners with respect to their base positions, over the course of a simulation. The specific combination of positioner design, dimensions and relative spacing/overlap is imprinted within these patterns. The high and low resolution target catalogues have different sky densities and relative priorities, these effects also imprint themselves on the scatter plots (c.f. Table 1). We supply one file per simulation. Each file has one page where the distribution of fiber offsets is illustrated first with a scatter plot, and then as a binned density plot (binned with 0.5x0.5mm pixels). A separate panel is shown for each combination of fiber resolution (High or Low res) and moon phase (Dark+Grey or Bright time). The colour scale is linear and indicates the relative density per pixel. Only fiber offsets from the first tile in each field are shown. The number of fiber offsets in each plot is limited to 100000. Only positioners in the central part of the FOV are shown, i.e. only those positioners with a full set of neighbouring positioners. The locations of neighbouring positioners are indicated with black boxes. If the positioners have a minimum patrol radius, then this is indicated with a grey disc. Note, when in the 'resting' position (i.e. before it has been assigned to a target), the PotzPoz positioner measures 5.6mmx2mm with the long dimension parallel to the y-axis in these plots. This is imprinted as a 'zone of avoidance' around each of the neighbouring positioner locations. Note that high resolution objects are almost always allocated fibers before low resolution objects. Therefore high resolution positioners almost always encounter neighbouring low resolution positioners in their resting positions. When a high resolution object has been marked as 'completed', then it is moved down the priority queue but not totally ignored.

Hence in some rare instances a high-resolution positioner can be assigned after neighbouring low resolution positioners have been moved from their resting positions. Note that the '1in6' and '1in3' patterns for high resolution fiber placement are not isotropic, and so the fiber radial scatter plots contain a superposition of the patterns caused by all possible relative fiber-fiber spacings.

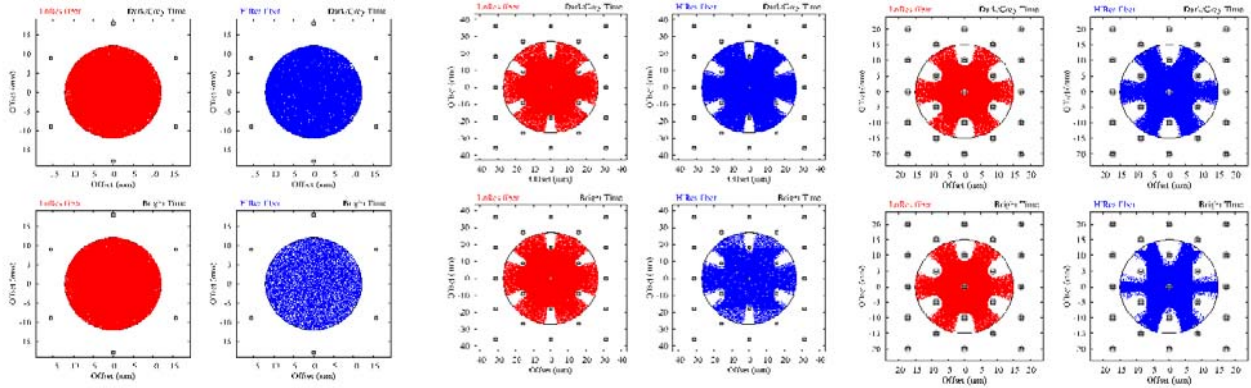

Figure 3:. Fiber positioner offset density for the NTT and VISTA telescope and the MuPoz and Potzpos positioner. The left four plots refer to the offset densities for NTT+MuPoz, the middle panel is for NTT+PotzPoz, and the right four plots refer to the simulation output for VISAT+PotzPoz. See the text for details.

**3.4.2 Numerical values of the sky area accessible by each positioner for the different simulations**

Table 1 gives the sky area accessed per positioner, determined from simuations of the 4MOST sky survey, for several system configurations.

Table 3:.This table lists the sky areas accessible per positioner for a number of simulated instrument configurations. The columns give the telescope (which determines the plate scale, 0.1041 mmarcsec$^{-1}$ for NTT and 0.0574 mmarcsec$^{-1}$ for VISTA), the positioner concept, the instrumental field of view (FOV, in deg$^2$), the number of fibers (nfibs), and the ratio of the number of high versus low resolution fibers (hilo). The last three columns give the areas (in arcmin$^2$) of the nominal circlular area patrolled by each positioner (APD), and the average area accessed by high resolution (AHR), and low resolution (ALR) positioners.

| *Telescope* | **Positioner** | **FOV** | **nfibs** | **hilo** | **APD** | **AHR** | **ALR** |
|---|---|---|---|---|---|---|---|
| *NTT* | MuPoz 3 | 3.0 | 1500 | 1:6 | 11.93 | 11.55 | 11.56 |
| *NTT* | PotzPoz | 3.0 | 1500 | 1:6 | 59.56 | 47.97 | 48.97 |
| *VISTA* | PotzPoz | 3.0 | 1500 | 1:6 | 61.08 | 43.15 | 44.76 |
| *NTT* | MuPoz, | 4.25 | 1500 | 1:6 | 16.94 | 16.37 | 16.51 |
| *NTT* | PotzPoz | 4.25 | 3000 | 1:9 | 42.13 | 32.36 | 34.16 |
| *VISTA* | PotzPoz | 6.1 | 3000 | 1:9 | 61.08 | 39.77 | 44.63 |
| *VISTA* | PotzPoz | 4.25 | 2250 | 1:9 | 58.55 | 37.17 | 42.39 |
| *NTT* | PotzPoz | 3.0 | 3000 | 1:9 | 31.27 | 23.74 | 24.56 |
| *VISTA* | PotzPoz | 3.0 | 3000 | 1:9 | 32.08 | 18.74 | 19.76 |
| *NTT* | MuPoz | 4.25 | 2250 | 1:9 | 11.93 | 11.52 | 11.56 |
| *VISTA* | PotzPoz | 4.25 | 1500 | 1:6 | 87.30 | 62.46 | 67.84 |
| *VISTA* | PotzPoz | 6.1 | 1500 | 1:6 | 214.38 | 132.66 | 150.21 |
| *VISTA* | PotzPoz | 4.25 | 2250 | 1:3 | 58.55 | 38.42 | 2.54 |

# 4. 4FS FRACTION OF INPUT OBJECTS TO ASSIGNED FIBERS

Figures. 5, 6, and 7 illustrate, in a graphical way, the summary statistics internally generated by the 4FS OpSim for each simulation that has been carried out. The plots show the fraction of input objects that have been assigned a fiber in at least one tile of the simulated survey. Note that this is not the same measure as the fraction of input objects that were 'successfully' observed reported by the 4FS DQCT. However, the allocated fractions reported here do provide strong upper bounds on the successfully observed fractions. The survey footprint is defined as the locus of all points on the sky that lie within the hexagonal bounds of at least one field in which at least one tile was executed. Due to the FOV shape, the survey footprint has slightly 'ragged' Northern and Southern edges. Therefore, some points of the footprint extend slightly outside the nominal declination limits of the survey. The plots illustrate the relative ability of each combination of telescope+positioner+FOV+number of fibers+high/low resolution fiber pattern that has been simulated. Each of these pages shows the results for one DRS. The red bars show the fraction of objects within the survey footprint that were allocated a fiber in at least one tile. The blue bars show the fraction of all input objects (within the declination limits -70<Dec<+20 deg) that were allocated a fiber in at least one tile. The total number of objects within the -70<Dec<+20 deg range is given in the top right hand corner of the plot. The red and blue bars should be read off against the left hand y-axis scale. The green bars show the number of objects that were allocated a fiber in each simulation, with a scale that should be read off the right hand y-axis. Note that the blue bar is simply the number of sources allocated fibers (indicated by the green bar) divided by the total number of objects, indicated in the top right hand corner of each plot.

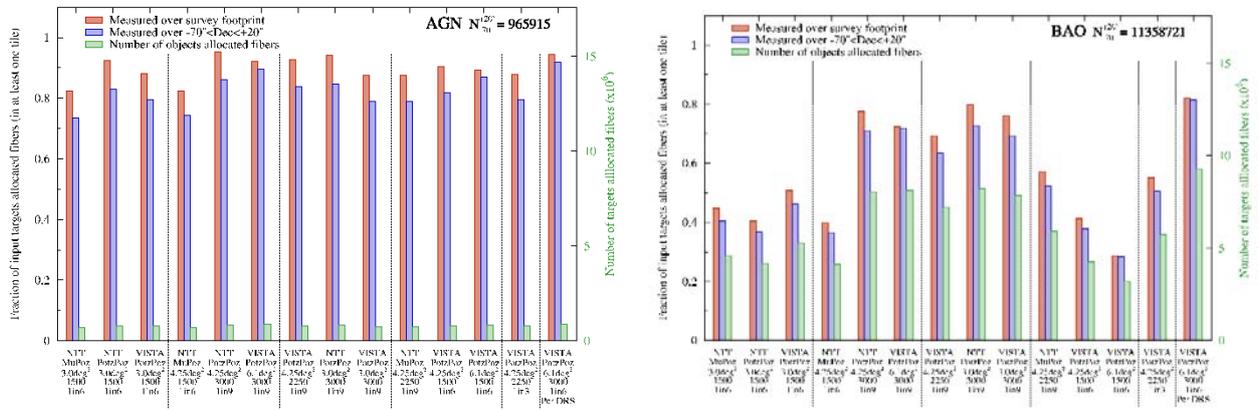

Figure 5: Fraction of input objects that have been assigned a fiber in at least one tile of the simulated survey. Note that this is not the same measure as the fraction of input objects that were 'successfully' observed reported by the 4FS DQCT. The results for the simulation for the AGN and the BAO DRS are shown.

Figure 6: The results for the simulation for the Galactic Halo Low Resolution and the Galactic Disk Low Resolution DRS are shown.

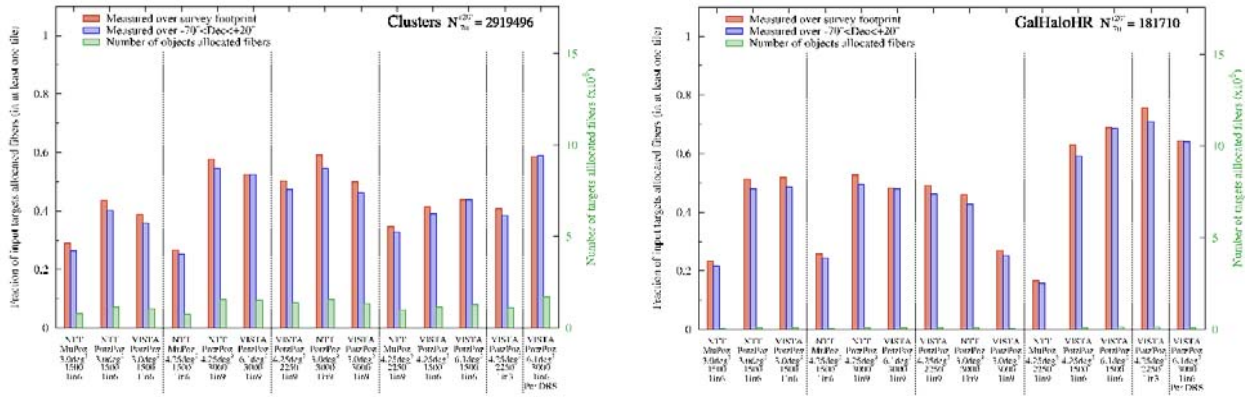

Figure 7: The results for the simulations of the Clusters and the Galactic Halo High Resolution DRS are shown.

## 5. SUMMARY

We have presented the initial results from the 4MOST Facility simulator. Using these simulations we can easily differentiate between several different conceptual instrument configurations in their ability to complete design reference sky surveys, to and achieve a set of baseline science goals. We emphasize the value of having a functioning facility simulator very early in the conceptual design phase of a large and complex instrument such as 4MOST. While the Operation Simulation and the Targeting algorithm are only a part of the 4FS system, our simulations have been used to inform the 4MOST Telescope Trade Off Committee. The Trade Off Committee has recommend the VISTA telescope as the location of the 4MOST instrument and survey in early May 2012. On May 30, 2012, ESO finally selected VISTA as the telescope for further studies.

## ACKNOWLEDGMENTS

TB and TD are grateful to Hans Böhringer, the 4MOST manager at MPE, and to Jakob Walcher, the Instrument Scientist of 4MOST, for a fruitful and very constructive collaboration. The authors especially acknowledge the efforts of the PI of 4MOST, Roelof de Jong, to make 4MOST to a success for the whole collaboration.

## REFERENCES

[1] R. H. Hardin, N. J. A. Sloane and W. D. Smith, Tables of spherical codes with icosahedral symmetry, published electronically at http://www.research.att.com/~njas/icosahedral.codes/

Updated 1 March 2012